\documentclass{cjaa}                   

\usepackage{graphicx}                   

\usepackage{threeparttable}
\begin{document}

   \title{A GRB Follow-up System at the Xinglong Observatory
   and Detection of the High-Redshift GRB 060927
   \footnote{Supported by the National Natural Science Foundation of China.}}

   \volnopage{Vol. 8 (2008) No.6, 693--699}      
   \setcounter{page}{1}          
   \baselineskip=3.7mm             

   \setcounter{page}{1}          

\author{Wei-Kang Zheng\inst{1,2}, Jin-Song Deng\inst{1}, Meng Zhai\inst{1,2}, Li-Ping Xin\inst{1,2},
    Yu-Lei Qiu\inst{1}, Jing Wang\inst{1}, Xiao-Meng Lu\inst{1,2}, Jian-Yan Wei\inst{1}, Jing-Yao Hu\inst{1}
      }

   \institute{National Astronomical Observatories, Chinese Academy of Sciences,
             Beijing 100012, China;\\
             \hspace{0.3cm}{\it zwk@bao.ac.cn}
             \and
         Graduate School of Chinese Academy of Science, Beijing 100049, China
          }

   \date{Received 2007 December 30; accepted 2008 March 27}

   \abstract{
A gamma-ray burst (GRB) optical photometric follow-up system at
the Xinglong Observatory of National Astronomical Observatories of
China (NAOC) has been constructed. It uses the 0.8-m Tsinghua-NAOC
Telescope (TNT) and the 1-m EST telescope, and can automatically respond
to GRB Coordinates Network (GCN) alerts. Both telescopes slew relatively
fast, being able to point to a new target field within $\sim 1$~min
upon a request. Whenever available, the 2.16-m NAOC telescope is also
used. In 2006, the system responded to 15 GRBs and detected seven early
afterglows. In 2007, six GRBs have been detected among 18 follow-up
observations. TNT observations of the second most distant GRB 060927
($z=5.5$) are shown, which started as early as 91~s after the GRB
trigger. The afterglow was detected in the combined image of first
$19\times20$~s unfiltered exposures. This GRB follow-up system has
joined the East-Asia GRB Follow-up Observation Network (EAFON).
   }

   \authorrunning{W. K. Zheng et al.}            
   \titlerunning{A GRB Follow-up System at Xinglong and GRB 060927}  

   \maketitle

\section{Introduction}           
\label{sect:int}
Gamma-Ray bursts (GRBs) are short-lived, intense flashes of gamma
rays from space, lasting from a few milliseconds to many minutes
(see Zhang \& M\'{e}sz\'{a}ros 2004 for a review). They were
first discovered by accident by the US military VELA satellites in the
late 1960's (Klebesadel et al. 1973), but their origin was an
enigma to astronomers for nearly 30 years. The breakthrough came in
1997 when for the first time the X-ray and optical counterparts
(i.e. afterglows and host galaxies) of a GRB were found (van Paradijs et al.
1997). This in turn enabled the distance to be established as
``cosmological'', with a typical redshift of $z\sim1-3$ (Jakobsson
et al. 2006). The observations of the $Swift$ satellite since its
launch in the late 2004 (Gehrels et al. 2004), in particular in the
soft X-ray band, have provided considerable new insight into the
nature of GRBs (e.g., Yu \& Huang 2007; Gao \& Fan 2006), as well as
raising many new problems. A comprehensive review of new progress in
the $Swift$ era can be found in Zhang (2007).

Fast follow-up observations in the optical are still of paramount
importance for the understanding of GRBs. Now, optical
afterglows in early phase contain fruitful information on GRB
physics (e.g., Fan et al. 2002; Zhang et al. 2003; Yan et al. 2007),
compared with the later self-similar Sedov stage. On the other hand,
GRB afterglows are optical transients, whose brightness declines
roughly as a power law with a typical index of $\sim1$ (Wu et al.
2004; Liang \& Zhang 2006). For example, a ``bright'' afterglow of
$\sim 15$~mag in 100~s may already become as faint as $\sim
19$~mag just 1 hour later.

Since the late 1990's, many fast-responding GRB optical follow-up
systems have been in operation around the world attached to small or
middle-size telescopes (ROTSE-III, TAROT, RAPTOR, PROMPT, KAIT,
REM, Liverpool \& Faulkes, etc., just to name a few). However, by 2004
such a facility was still conspicuously absent in China despite
the vast longitudinal and latitudinal range the country spans. Then
for a GRB that happens to occur in the night sky, the golden chance to
catch its early afterglow could be missed.

In this paper, we report the GRB optical photometric follow-up
system at the Xinglong Observatory of National Astronomical
Observatories of China (NAOC) that we have constructed since 2004,
concentrating mainly on the second routine run that started in
early 2006, although the first observation was made as early as
in 2004 April (Qiu \& Hu 2004). Our telescope system, observing
program and strategy are described in Sections 2 and 3, respectively.
The performance of our system in 2006 and 2007 is summarized in
Section 4. The detection of the optical afterglow of the very
high-redshift GRB 060927 ($z=5.5$) is presented in Section 5,
which demonstrates a good capability for GRB investigations.

\section{The Telescope System} \label{sect:tel}

Our telescopes are located at one of the major optical astronomy
sites in East Asia, the Xinglong Observatory of NAOC. The site is
about 170~km to the northeast of Beijing and about 900~m above sea
level, at longitude $7^{\rm h}50^{\rm m}18^{\rm s}$ east and
latitude $40^\circ23'36''$ north. On average there are $\sim 240-260$
spectroscopic nights and $\sim 100-120$ photometric nights every year.

The development of the system started with the 80-cm Tsinghua-NAOC
Telescope (TNT) in early 2004. This is an equatorial-mounted Cassegrain
system with a focal ratio of $f/10$, made by AstroOptik, founded
by Tsinghua University in 2002 and jointly operated with NAOC.
TNT is our primary instrument for GRB photometric follow-up observations,
although it also serves more general purposes like the monitoring of
supernovae, blazars, and variable stars (e.g. Wu et al. 2005).

Since 2006 our system has been enlarged to include the 1-m EST telescope
after its installation was completed in 2005. This telescope, manufactured
by the EOS Technologies, is a Ritchey-Chr\'{e}tien system and is
alt-azimuth mounted. It features dual Nasmyth focal positions at
a focal ratio of $f/8$ with two field de-rotators.

The two telescopes are equipped with the same type of Princeton Instrument
$1340\times1300$ thin back-illuminated CCD, with pixel size of
$\sim 20~\mu$m and liquid-nitrogen cooled. The resulting field
of view, i.e. $\sim 11'\times11'$ at the aforementioned focal
ratios, well covers the typical $\sim 1'-4'$ location error circle
of the $Swift$ Burst Alert Telescope (BAT), the main provider of
real-time GRB alerts since 2004. Both CCD cameras are covered with
standard Johnson-Cousin $UBVRI$ filters made by Custom Scientific,
but we also took unfiltered exposures.

The telescopes can respond quickly to a request for GRB
observations. The TNT slews at a speed of $\sim 2^\circ$ per second,
while the EST is about two times faster. The traditional rotating
domes that house them also move fast. A software system has been
developed that can automatically take control of the telescope and
CCD camera immediately upon receiving a GRB Coordinates Network
(GCN) alert via a socket connection.

The 2.16-m NAOC optical telescope may also be used if the Beijing
Faint Object Spectrograph and Camera (BFOSC) is mounted at the time.
The camera uses a Loral Lick 3 thin $2048\times2048$ 15-$\mu$m CCD
that provides a field of view of $\sim 10'\times10'$ for direct
imaging. Standard Jhonson-Cousin $UBVRI$ filters are also fitted.
However, this relatively old telescope cannot be used in fast
response: it was only used for later time faint GRB afterglows.

\section{The Observing Program and Strategy} \label{sect:str}

Our GRB follow-up system and observation program, to a large
extent automatic, are outlined in Figure 1. Automation is required
in order to catch the fast-fading GRB afterglows early on. It has
been implemented in most systems dedicated to GRB follow-up projects,
pioneered by the French TAROT (B\"{o}er et al. 1999).

\begin{figure}[!hbp]
\centering
\includegraphics[width=0.9\textwidth]{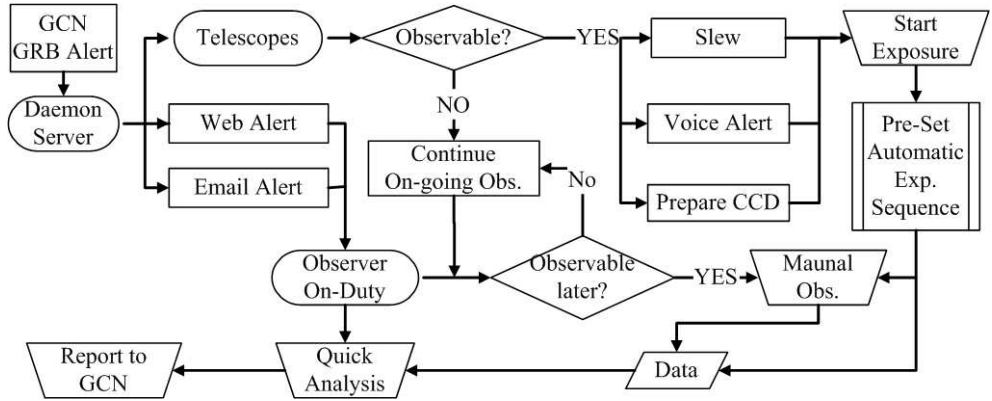}
\caption{Flow chart of the GRB follow-up system. {\em Trapeziums} mark
manual operations, and {\em rectangles} mark automatic ones.}
\end{figure}

In our system, a daemon program receives and reduces the GRB alert
messages that are distributed by GCN through the internet. It
continuously communicates with GCN through a socket interface.
Although GRB triggers of various satellites are available,
currently the program only lets through those of the $Swift$,
whose BAT detects the most GRBs with locations and provides
on-board location error circles as small as $\sim 1'-4'$ to the
ground within just 20~s (Barthelmy et al. 2005). The trigger is
then broadcast to our system connected in the local network, which
is sent to the team members via email and recorded in a dedicated
web site.

Once the system obtains the alert, a software will check the GRB
coordinates. If the GRB is observable at that time, the software
will immediately point the telescope at it and prepare the CCD
camera for GRB imaging, overriding any on-going observation. A
voice alert is also generated to inform the night assistant to
start the pre-set sequence of automatic CCD exposures. This human
intervention means no more than a hit of the keyboard in most
cases and can be avoided with further automation.

An observer on-duty is activated after receiving the alert email
or checking the web site, and he/she shall make a quick analysis of
the observed data to look for an optical counterpart, and
circulate the results to GCN. It is also his/her responsibility to
plan any manual observation, if the GRB is to rise above the
horizon later that night, or after the pre-set exposure sequence
is completed.

The pre-set sequence starts with 20$\times$20~s unfiltered
automatic exposures, within a total of $450-500$~s. The
unfiltered, i.e. the white band, is used in order to have the best
sensitivity, since it is all-important to have early detection of
an optical counterpart. With a 20~s white-band exposure, the TNT
can reach a limiting magnitude of $\sim 18-19$, thus sensitive to
a large fraction of afterglows within $\sim 1000$~s after GRB
(Guidorzi et al. 2006). The CCD operates in high-speed mode to
minimize the read-out time to only $\sim 1.8$~s per image, in an
effort to keep the details of temporal variability of the afterglow.

Then 20$\times$60~s $R$-band automatic exposures are carried on.
The CCD is switched to low-speed mode to reduce the noise, with
a read-out time of $\sim 18$~s per image. The increased exposure
time matches the fast-decaying brightness of GRB afterglows. After
the pre-set sequence, only exposures of 300~s are taken and the
observer on-duty will decide when to change the filter and when to
terminate the observation.

The above program and strategy are for the TNT. For the EST, the GRB
alert with coordinates is displayed on the telescope-control computer
in real time. The observer on-duty will interrupt an on-going
observation to let the EST follow up a GRB trigger. Nevertheless,
we continue to update the control software of the EST to make its
GRB observations automatic and simultaneous with the TNT but in a
different photometric band.

Flat-field and bias frames are routinely obtained by the
night assistant during dusk or dawn. No dark frame is needed since
our CCDs are sufficiently cooled. For any GRB afterglow detected,
the data reduction will be done after the reference stars in the
GRB field have been calibrated in a good photometric night.

\section{The Observation Performance} \label{sect:per}

The first routine run of our system was mainly in 2004, mainly
for GRBs localized by HETE-2. However, only one afterglow, that of
GRB 041006 about 0.1 day after the trigger, was detected (Urata et
al. 2007a), owing to the relatively low event rate and relatively
large error circle ($\sim 30'$ by the HETE-2 Wide-Field X-ray
Monitor, Shirasaki et al. 2003). The system was unfortunately
disabled from running automatically in the first year of the $Swift$
era by an accidental loss of socket connection to GCN.

The second routine run\footnotemark{}\footnotetext{Real-time updated
log can be found at: http://www.xinglong-naoc.org/grb/index.html}
began in 2006 February after the socket connection had been recovered
and the system refurbished. In 2006 and 2007, 13 afterglows were
detected among the 33 bursts observed\footnotemark{}\footnotetext{
GRB 060124 was observed under the request of our collaborators in
the East-Asia GRB Follow-up Observation Network (EAFON):
http://cosmic.riken.go.jp/grb/eafon/index.html}. All the bursts were
localized by $Swift$, except GRB 060930 and GRB 070125.

For 17 of the $Swift$ GRBs, our observations were automatically
triggered, and started within $\sim 1000$~s of their first BAT
detections, which largely matches the expectations. Simple statistics
show that $\sim 20\%$ of the $\sim 100$ $Swift$ GRBs each year were
discovered by BAT during nighttime hours with altitudes
$>20^\circ$ at our observatory\footnotemark{}\footnotetext{
http://swift.gsfc.nasa.gov/docs/swift/archive/grb\_table/}.
The expected number would be $\sim 25$, allowing for the $\sim
65\%$ of observable nights. The small residual difference may be
accounted for by the a few non-operational days of our telescopes
caused by hardware failures.

The detection rate is the highest for the six GRBs whose observation
start-time was less than 2~min after the space $\gamma$-ray
instrument was triggered, which is $\sim 70\%$. It deteriorates to
$\sim 50\%$ for the whole automatic sample and further to $\sim
25\%$ for those of only relatively-late manual observations (i.e.
$>1$~hr). Note that $\sim 65\%$ of our undetected GRBs have no
afterglow identified by other optical telescopes either.

Among the detected afterglows, we have the earliest photometry for
GRB 060912A and GRB 061110A, even before the onboard
Ultra-Violet/Optical Telescope (UVOT) of  $Swift$. In the cases of
GRB 060323 and GRB 070518, our detections seems to preced any other
ground telescopes. We have detected the very high-redshift GRB
060927, as shown in the next section. We also note that had it
not been our automatic observations taken between UVOT and other
ground telescopes, that of GRB 060323 would have been accepted as
``one of the faintest afterglows ever discovered'' (Kann et al.
2006). Detailed analysis of the afterglows will be presented
elsewhere, although preliminary results of GRB 060124, GRB 060218
and GRB 060323 have been shown in Deng et al. (2006).

\section{The Very High-Redshift GRB 060927}  \label{sect:grb}

GRBs have long been believed to be promising powerful probes of the
early universe thanks to their extreme brightness (e.g. Lamb \&
Reichart 2000). In fact, the most distant GRB 050904 has already
been used to explore the universe's reionization epoch (Totani
et al. 2006). However, so far there are still only four GRBs with a
measured redshift of $z>5$, which are GRB 050814 ($z\sim 5.3$), GRB
050904 ($z=6.3$), GRB 060522 ($z=5.1$) and GRB 060927 ($z=5.5$), all
detected by
$Swift$\footnotemark{}\footnotetext{http://www.mpe.mpg.de/$^{\sim}$jcg/grbgen.html}.

GRB 060927 triggered the $Swift$ BAT at 14:07:35 UT on 2006
September 27. An optical counterpart was detected by ROTSE-III in
$<20$~s (Schaefer et al. 2006), and was soon confirmed by other
telescopes including the TNT (Zhai et al. 2006). A redshift of
$z=5.6$ was found by Fynbo et al. (2006) through VLT spectroscopy,
making it the second most distant GRB. The value was later updated
to 5.47 (Ruiz-Velasco et al. 2007).

\begin{threeparttable}[!hbp]
  \caption[]{Observation Log of 2006 and 2007.}
  \label{Tab:ObsLog}
  \begin{center}\begin{tabular}{lccccc}
  \hline\noalign{\smallskip}
 GRB & Time start & Filters & Telescopes & Detection? & GCN Circ.\\
  \hline\noalign{\smallskip}
071112C & 113s  & W,R,V & TNT,EST & YES & 7063\\
071101  & 94s   & W,R & TNT & NO  & 7036\\
071028A & 203s  &  W,R,V  & TNT,EST & NO  & 7015\\
071025  & 6.3h  &  I  & TNT & NO  & 6999\\
071021  & 814s  & W,R & TNT & NO  & 6962\\
071020  & 10.1h &  R  & TNT & YES & 6956\\
071018  & 10.44h&  R  & TNT & NO & 6936,6965\\
071013  & 289s  & W,R & TNT & NO & 6908\\
071011  & 353s  &  W,R,V& TNT,EST & YES & 6904\\
071010B & 23.31h&  V  &  EST &  YES&  6924\\
070810B & 300s  &  W,R & TNT &  NO &  6747\\
070808  & 115s  &  W,R & TNT &  NO &  6722\\
070529  & 2.03h &  R,V & TNT &  NO &  6467\\
070520A & 579s  &  R  &  EST &  NO &  6424\\
070518  & 512s  & W,R,V,I &  TNT,EST & YES & 6416\\
070406  & 19.86h&  W  &  TNT &  NO  & 6250\\
070129  & 11.98h&  R  &  TNT &  NO  & 6057\\
070125  & 27.6h & R,V &  TNT,EST &  YES & 6035\\
061222A & 9.97h &  R  &  2.16m &  NO  & 5976\\
061110A & 76s   & W,R &  TNT &  YES & 5798\\
060930  & 2.74h &  R  &  TNT &  NO  & 5669\\
060927  & 91s   & W,R &  TNT &  YES & 5638\\
060923C & 215s  & W,R &  TNT &  NO  & 5596\\
060912A & 89s   & W,R &  TNT &  YES & 5560\\
060605  & 231s  & W,R &  TNT &  YES & 5230\\
060502B & 305s  & W,R &  TNT &  NO  & 5057\\
060428B & 3.54h &  I  &  TNT &  NO  & 5028\\
060427A & 4.79h &  R  &  TNT &  NO  & 5015\\
060403  & 6.07h &  R  &  TNT &  NO  & 4955\\
060323  & 540s  &  W  &  TNT &  YES & 4930\\
060223A & 5.42h &  W  &  TNT &  NO  & 4827\\
060218  & 7.20h &  W,R,V,I,B  &  TNT,EST,2.16m &  YES & 4802\\
060124  & 0.45h &  R,I  &  TNT &  YES & 4588\\
  \noalign{\smallskip}\hline
  \end{tabular}
  \end{center}
\end{threeparttable}
\\
\\

The TNT responded to this burst as early as 91~s after the BAT
trigger, or within 80~s of the receipt of the GCN alert. This
was only second to ROTSE-III. The exposure sequence was slightly
different from the routine one described in Section 3; it produced
$19\times20$~s white-band images followed by seven 60~s and
three 600~s $R$-band images. The observation was ended when the
observer on-duty judged that the afterglow had become too faint
to be detected.

The afterglow is identified as a $3.2\sigma$ source in the
combined image of first 19$\times$20~s exposures, as shown in
Figure 2. It can not be seen in any single-exposure image, nor in
the combined later $R$-band images. The image reduction (including
bias subtraction and flat-field correction) was performed using
standard IRAF\footnotemark{}\footnotetext{IRAF is distributed by
NOAO, which is operated by AURA, Inc., under cooperative agreement
with NSF.}. Differential aperture photometry was
performed using the APPHOT package in IRAF and using the $R$-band
magnitudes of reference stars, which were calibrated by observing
Landolt standards on 2006 December 21. The afterglow magnitudes
so obtained are listed in Table 2.

\begin{figure}[]
\centering
   \includegraphics[bb=0 0 504 372,width=0.7\textwidth]{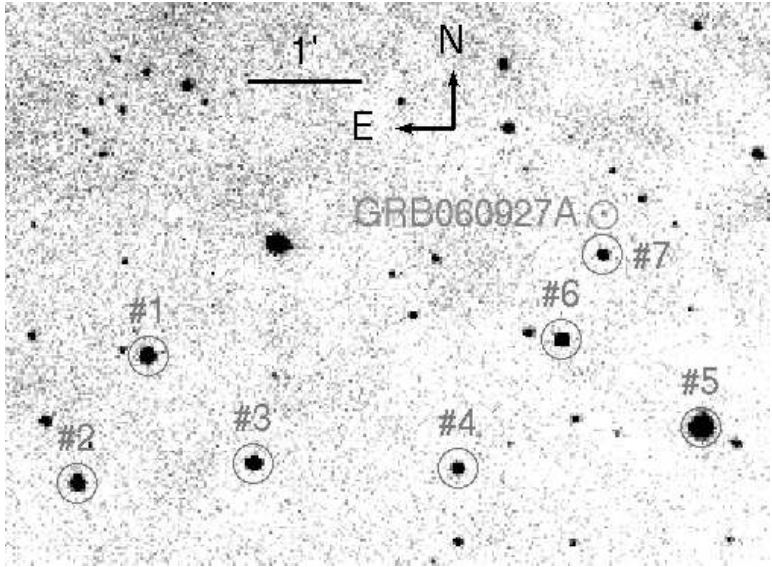}
   \caption{A combined TNT white-band image of GRB 060927. The GRB afterglow
   is indicated by the {\em small circle}, while the photometric reference
   stars by {\em large circles} labelled \#1 to \#7.}
\end{figure}

\begin{table}[!hbp]
  \caption[]{Data of GRB 060927 with the TNT}
  \begin{center}
  \begin{tabular}{ccccccc}
  \hline\noalign{\smallskip}
Time Start & Time End & Mean Time & Exp Time & Filter & Mag & Upper Limit\\
  \hline\noalign{\smallskip}
91s & 572s & 331.5s & 19 $\times$ 20s & White  & 19.67$\pm$0.33 & NO \\
626s & 1214s & 920s & 7 $\times$ 60s & R & $>$19.25 & YES \\
1434s & 2430s & 1932s & 3 $\times$ 300s & R  & $>$20.02 & YES \\
\hline\\
\end{tabular}
\end{center}
\end{table}

Caution must be exercised while comparing our white-band results with
those of ROTSE-III as published in Ruiz-Velasco et al. (2007).
Although the spectral response of an unfiltered CCD typically peaks
in the $R$ band, for such a high-$z$ object the Ly$\alpha$
blanketing absorption at $<8000{\rm\AA}$ makes the white band an
effective $I$ band. For the first two ROTSE-III white-band
measurements, both preliminary $R$-band calibrated results
(Schaefer et al. 2006) and final $I$-band calibrated results
(Ruiz-Velasco et al. 2007) were reported. The difference between
them is as large as $\sim 2$~mag. Assuming that this value is also
applicable to our data, the $\sim 19.7$~mag at $\sim 5.5$~min
in Table 2 would correspond to $\sim 17.7$~mag, consistent
with the $I$-band calibrated ROTSE-III result of $\sim 17.9$~mag
at $\sim 4.6$~min (Ruiz-Velasco et al. 2007).

\section{Conclusion and Future Plans} \label{sect:con}

Our GRB optical follow-up system ran efficiently in 2006 and 2007,
resulting in the automatic detection of eight very early afterglows
including that of the second most distant GRB. Our system has
advantage in terms of limiting magnitude over the dedicated,
ultra-fast, small-size telescopes that can slews at several tens of
degrees per second, similar to the 25-cm TAROT and 45-cm ROTSE-III
(Akerlof et al. 2003). Within the East-Asia GRB Follow-up
Observation Network (EAFON, Urata et al. 2005), our observations of
early-phase afterglows and the later but deeper detections made by
the Kiso 1.05~m and Lulin 1-m telescopes, are complementary to each
other (e.g., Huang et al. 2005, 2007; Urata et al. 2007b), since
a long time coverage of optical afterglows is very important for
understanding the nature of GRBs (e.g., Huang \& Cheng 2003; Urata et al. 2007c).

Further automation is planned in order to fully exploit the
capacity of the system, e.g. the fast slewing speed of the EST. A
three-channel camera based on CCD detectors is under development,
which will allow our telescope to perform simultaneous multi-color
photometry without loss of sensitivity and time resolution. We are
also developing a pipeline for real-time data analysis. In addition,
the CCD exposure sequence will be optimized regarding the use of
the white band, which was both a blessing (i.e. good sensitivity)
and a pain (i.e. no reliable calibration at all).

\begin{acknowledgements}
We are grateful to Yuji Urata and Lijin Huang for helping setting up
our GRB follow-up system, to Xiaofeng Wang and other THAC members
for their generosity in sharing telescope time, and to Scott
Barthelmy for providing GCN connection. This study has been
supported by the NSFC (No. 10673014).
\end{acknowledgements}

\label{lastpage}


\begin{thebibliography}{99}

  \bibitem[2003]{} Akerlof C. W., Kehoe R. L., McKay, T. A. et al., 2003, \pasp, 115, 132

  \bibitem[2005]{} Barthelmy S. D. et al., 2005, \ssr, 120. 143

  \bibitem[1999]{} B\"{o}er M. et al., 1999, \aaps, 138, 579

  \bibitem[2006]{} Deng J. et al., 2006, Il Nuovo Cimento B, 121, 1469

  \bibitem[2002]{} Fan Y.Z., Dai Z.G., Huang Y. F. et al., 2002, Chin. J. Astron. Astrophys.(ChJAA), 2, 449

  \bibitem[2006]{} Fynbo J. P. U. et al., 2006, GCN Circ., 5651

  \bibitem[2006]{} Gao W.H., Fan Y.Z., 2006, Chin. J. Astron. Astrophys.(ChJAA), 6, 513

  \bibitem[2004]{} Gehrels N. et al., 2004, \apj, 611, 1005

  \bibitem[2006]{} Guidorzi C. et al., 2006, \pasp, 118, 288

  \bibitem[2005]{} Huang K.Y. et al., 2005, \apj, 628, L93

  \bibitem[2007]{} Huang K.Y. et al., 2007, \apj, 654, L25

  \bibitem[2003]{} Huang Y.F., Cheng K.S., 2003, MNRAS, 341, 263

  \bibitem[2006]{} Jakobsson P. et al., 2006, \aap, 447, 897

  \bibitem[2006]{} Kann D. A. et al. 2006, GCN Circ., 4913

  \bibitem[1973]{} Klebesadel R., Strong I., Olson R., 1973, \apj, 182, L85

  \bibitem[2000]{} Lamb D. Q., Reichart D. E., 2000, \apj, 536, L1

  \bibitem[2006]{} Liang E., Zhang B., 2006, \apj, 638, L67

  \bibitem[1997]{} van Paradijs J. et al., 1997, \nat, 386,686

  \bibitem[2004]{} Qiu Y., Hu J., 2004, GCN Circ. 2581

  \bibitem[2007]{} Ruiz-Velasco A. E. et al., 2007, \apj, 669, 1

  \bibitem[2006]{} Schaefer B. E., Yost S. A., Yuan F., 2006, GCN Circ., 5629

  \bibitem[2003]{} Shirasaki Y. et al., 2003, \pasj, 55, 1033

  \bibitem[2006]{} Totani T. et al., 2006, \pasj, 58, 485

  \bibitem[2005]{} Urata Y. et al., 2005, Il Nuovo Cimento C, 28, 775

  \bibitem[2007]{} Urata Y. et al., 2007a, \apj, 655, L81

  \bibitem[2007]{} Urata Y. et al., 2007b, \pasj, 59, L29

  \bibitem[2007]{} Urata Y. et al., 2007c, \apj, 668, L95

  \bibitem[2005]{} Wu C. et al., 2005, \aj, 130, 1640

  \bibitem[2004]{} Wu X.F., Dai Z. G., Huang, Y. F. et al., 2004, Chin. J. Astron. Astrophys.(ChJAA), 4, 455

  \bibitem[2007]{} Yan T., Wei D.M., Fan Y.Z., 2007, Chin. J. Astron. Astrophys.(ChJAA), 7, 777

  \bibitem[2007]{} Yu Y., Huang Y.F., 2007, Chin. J. Astron. Astrophys.(ChJAA), 7, 669

  \bibitem[2006]{} Zhai M. et al., 2006, GCN Circ., 5638

  \bibitem[2007]{} Zhang B., 2007, Chin. J. Astron. Astrophys.(ChJAA), 7, 1

  \bibitem[2003]{} Zhang B., Kobayashi S., M\'{e}sz\'{a}ros P., 2003, ApJ, 595, 950

  \bibitem[2004]{} Zhang B., M\'{e}sz\'{a}ros P., 2004, IJMPA, 19, 2385

\end{thebibliography}
\end{document}